# In-plane selective excitation of arbitrary vibration modes using thickness-shear ($d_{15}$) piezoelectric transducers

Hao Qiu[1], Faxin Li[1,2,a]

[1]LTCS, Department of Mechanics and Engineering Science, College of Engineering, Peking University, Beijing, 100871, China

[2]HEDPS, Center for Applied Physics and Technology, Peking University

**Abstract：**

Experimental modal analysis (EMA) is of great importance for the dynamic characterization of structures. Existing methods typically employ out-of-plane forces for excitation and measure the acceleration or strain for modal analysis. However, these methods encountered difficulties in some cases. In this work, we proposed an in-plane excitation method based on thickness-shear ($d_{15}$) piezoelectric transducers. Through the combination of distributed $d_{15}$ PZT strips, arbitrary vibration modes can be selectively excited in a wide frequency range. Both simulations and experiments were conducted and the results validated the proposed method. Specifically, bending, torsional, and longitudinal vibration modes of a rectangular bar were selectively excited. Torsional modes of a shaft were excited without the aid of brackets and bending modes of a circular plate were excited with actuators placed at nodal lines. Furthermore, the electromechanical impedance of the PZT-structure system was measured from which the natural frequency and quality factor were directly extracted. Due to its simplicity and flexibility, the proposed vibration excitation method is expected to be widely used in near future.

[a] Author to whom all correspondence should be addressed, Email: lifaxin@pku.edu.cn

# 1. Introduction

Experimental modal analysis (EMA) is one of the most important techniques for the dynamic characterization of structural systems [1]. In the past few decades, versatile vibration testing methods have been proposed, among which two of the most well-known are the impact hammer method [2] and the electrodynamic shaker method [3]. These two methods have achieved great successes in the EMAs of large structures. However, they encountered difficulties in dealing with small structures in which some vibration modes may not be easy to excite. To solve this problem, new vibration excitation methods have been proposed including non-contact and contact techniques.

The existing non-contact methods include the laser pulse method [4], the pressurized air method [5], the electromagnetic acoustic method [6], the ultrasound radiation method [7], and the eddy current method [8]. Since there is no contact between the actuator and the structure, these methods are especially suitable for the EMAs of small or micro structures. The main drawback of these methods is that they are typically more complex in contrast with the contact methods. The arrangement and calibration of the non-contact actuators usually take a long time and the testing repeatability is difficult to maintain. Moreover, the energy conversion efficiency of the non-contact methods is usually low and high power excitation is required.

Compared with the non-contact methods, the contact methods are more reliable and repeatable and had been widely used in researches and industries. Contact methods inevitably bring additional mass and stiffness which leads to measurement errors. However, this will not be a problem if the transducers are properly chosen and the errors are neglectable in most cases. Among all the contact-type transducers, piezoelectric transducers are the most frequently used ones because of their peculiar electromechanical coupling, quick response, and compact size. The existing piezoelectric transducers for EMA are typically based on the expansion ($d_{31}$ or $d_{33}$) mode [9–11]. When this type of piezoelectric transducers/wafer is bonded onto the structure with an applied voltage, both the out-of-plane forces and multidirectional in-plane forces will be generated on the structure surface. That is, a $d_{31}/d_{33}$ mode piezoelectric transducer behaves like a small size shaker.

However, it should be noted that the existing excitation methods may encounter problems when specific vibration modes are to be tested. For example, cylinders are widely used in shaft systems and their torsional vibration characteristics are highly concerned. However, currently, the excitation of torsional modes has to be aided with a specially designed bracket [12]. If an actuator that produces only unidirectional in-plane forces can be designed, torques can then be easily obtained by placing the actuators at the end of the circular bar. Actually, some piezoelectric transducers which have this feature have been proposed and studied in the guided waves area in the past few years [13–15]. Their effectiveness in exciting the torsional vibration of a cylinder has also been confirmed recently [16].

In this work, an in-plane excitation method that can be used for the excitation of arbitrary vibration modes was proposed based on distributed thickness-shear ($d_{15}$) mode piezoelectric. Unlike conventional EMAs [17] in which extra sensors are often required besides the actuators, in the proposed method, the piezoelectric strips can serve as both sensors and actuators. Only an impedance analyzer is required to measure the system's electromechanical impedance which reduces the testing complexity. Simulations were conducted on the selective mode excitation of a rectangular bar, a cylinder, and a circular plate. The rectangular bar was experimentally tested and the torsional, bending, and longitudinal vibration modes were selectively excited. The effectiveness of the impedance-based EMA was also experimentally validated by the velocity-based EMA using a 2D laser Doppler vibrometer.

## 2. Principle of the in-plane excitation method

2.1 Basic deformation modes of PZT ceramics

To illustrate the principle of the in-plane excitation method, the working principle of the piezoelectric transducers and their deformation modes are briefly reviewed here. Take the PZT (lead zirconate titanate) as an example, its constitutive equations can be written in a strain-charge form as:

$$\begin{aligned} \boldsymbol{S} &= \boldsymbol{s}^E \cdot \boldsymbol{T} + \boldsymbol{d}^t \cdot \boldsymbol{E} \\ \boldsymbol{D} &= \boldsymbol{d} \cdot \boldsymbol{T} + \boldsymbol{\varepsilon}^T \cdot \boldsymbol{E} \end{aligned} \quad (1)$$

where $\boldsymbol{S}$ is the strain tensor, $\boldsymbol{s}$ is the elastic compliance tensor under short-circuit conditions ($^E$ indicates constant electric field), $\boldsymbol{T}$ is the stress tensor, $\boldsymbol{d}$ is the piezoelectric coefficient tensor ($^t$ indicates transposition), $\boldsymbol{D}$ is the electric displacement vector, $\boldsymbol{\varepsilon}$ is the dielectric permittivity tensor ($^T$ indicates constant stress field) and $\boldsymbol{E}$ is the electric field vector.

In the absence of applied stress, the first equation in Eq. (1) can be written in matrix form as:

$$\begin{bmatrix} S_1 \\ S_2 \\ S_3 \\ S_4 \\ S_5 \\ S_6 \end{bmatrix} = \begin{bmatrix} 0 & 0 & d_{31} \\ 0 & 0 & d_{31} \\ 0 & 0 & d_{33} \\ 0 & d_{15} & 0 \\ d_{15} & 0 & 0 \\ 0 & 0 & 0 \end{bmatrix} \begin{bmatrix} E_1 \\ E_2 \\ E_3 \end{bmatrix} \quad (2)$$

where the subscripts have taken the Voigt notation and '3' denotes the direction of polarization. In the $\boldsymbol{d}$ matrix, $d_{33} > 0$, $d_{15} > 0$ and $d_{31} < 0$. Note that the poled PZT is transversely isotropic along the "3" direction, i.e., it is isotropic in the '1-2' plane.

From Eq. (2), it can be seen that when $E_3 \neq 0$ and $E_1 = E_2 = 0$, the strain components $S_1$, $S_2$, and $S_3$ will be generated and the PZT patch will extend (shrink) in the '3' direction and shrink (extend) in '1' and '2' directions. In the existing EMA method, piezoelectric actuators are designed based on this extension deformation mode. Under this mode, three normal strains will be generated together. So both the out-of-plane forces and multidirectional in-plane forces will be generated when the piezoelectric patch is bonded and used as an actuator. The "impure" forces can cause problems when natural frequencies of multiple vibration modes overlap and selectively excitation is needed.

By contrast, when $E_3 = 0$ but $E_1$ (or $E_2$) $\neq 0$, only $S_5$ (or $S_6$) will exist, which means a pure shear deformation in the '1-3' plane (or '2-3' plane) is generated. Using this deformation mode, we can apply an in-plane force to any part of the structure by bonding a $d_{15}$ piezoelectric strip onto the structure.

2.2 Selective vibration mode excitation based on distributed $d_{15}$ piezoelectric strips

Using different combinations of distributed $d_{15}$ mode piezoelectric strips, different vibration modes can be selectively excited. To clarify this idea, a rectangular bar is taken as an example. The commonly concerned vibration modes of a rectangular bar can be simply divided into three types: the bending modes, the longitudinal modes, and the torsional modes. To selectively excite these modes, only two piezoelectric strips are required as shown in Fig. 1(a). By properly setting the directions of the applied electric fields and polarization, as shown in Fig. 1(b-d), torques and forces can be generated and different vibration modes can thus be selectively excited. Similar arrangements can be used for structures of any other shape.

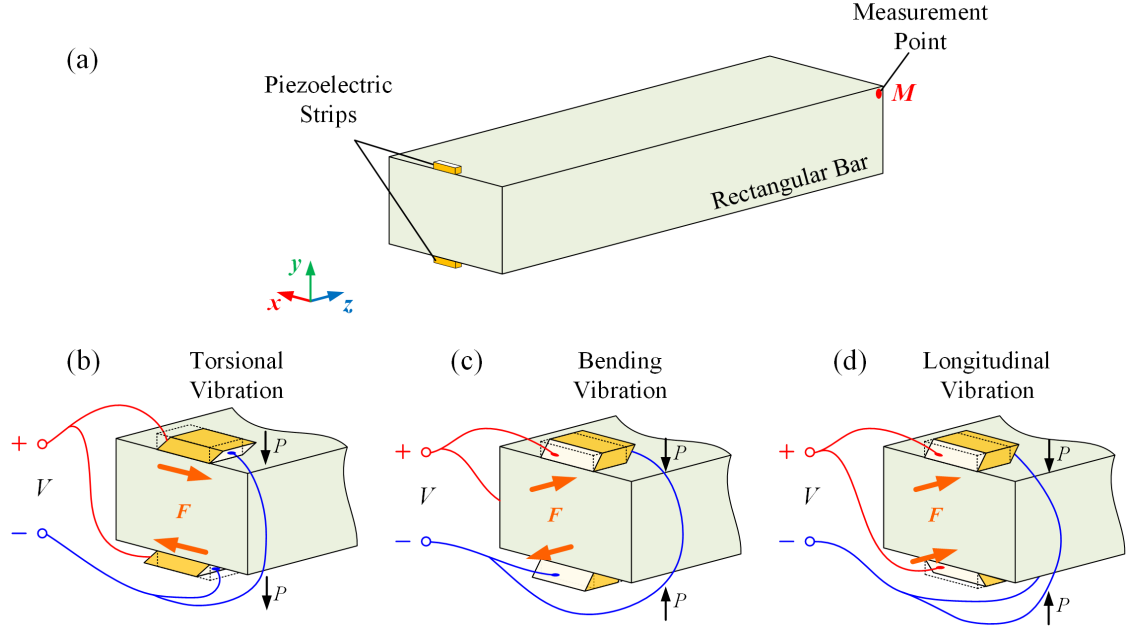

Fig. 1 Schematic layout of the in-plane excitation method for a rectangular bar. (a) The overall view; (b-d) The specific setups for selectively exciting torsional, bending, and longitudinal vibration modes, respectively. The polarization directions of the piezoelectric strips are denoted by the black arrows and the generated forces are denoted by orange arrows. White faces are silver electrodes.

2.3 Extracting the natural frequency and quality factor from the impedance curve

A promising advantage of this method is that by measuring the electromechanical impedance of the PZT-structure system using an impedance analyzer, both the natural frequency and qualify factor of the system can be directly obtained [18]. As shown in Fig. 2 and Eq.(3), the $n$th order natural frequency is the average of the resonance frequency ($f_r^{(n)}$) and aniti-resonance frequency ($f_a^{(n)}$) of the system and the quality factor is determined by the difference between them. It should be noted that typically the piezoelectric strip is much smaller than the structure. In this case, the natural frequency and qualify factor of the system can be regarded as that of the structure with the error typically less than 1%. Furthermore, in the case without joints, the quality factor of a structure made of single material is only determined by the internal friction of the material[19].

$$\begin{aligned} f_{resonance} &= \frac{f_a^{(n)}+f_r^{(n)}}{2} \\ Q &= \frac{f_a^{(n)}+f_r^{(n)}}{2\left(f_a^{(n)}-f_r^{(n)}\right)} \end{aligned} \tag{3}$$

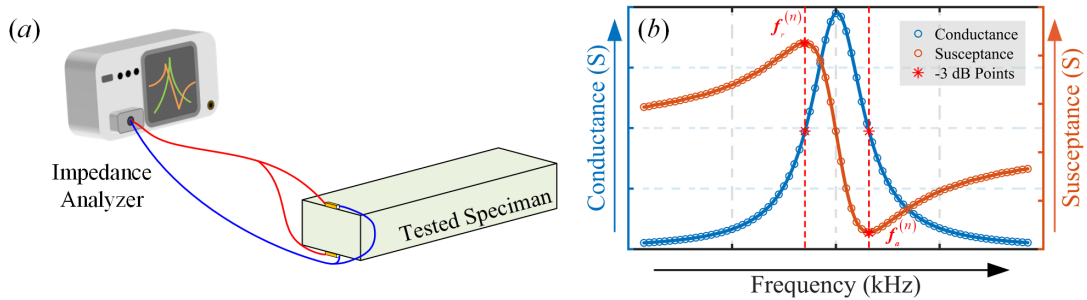

Fig. 2 Electromechanical impedance measurement and the natural frequency and quality factor calculation. (a) Schematic layout; (b) Typical conductance curve (blue circles) and the susceptance curve (orange, circles).

## 3. Simulations

### 3.1 Simulation setup

FE (Finite element) simulations were conducted to verify the effectiveness of this excitation method. The quality factor $Q$ or the internal friction of the system (material) was also taken into account during the simulations. For an isotropic material, the number of independent quality factors and elastic constants should be the same. So two types of $Q$ are defined in the simulations with the longitudinal quality factor $Q_E = 2000$ and the shear quality factor $Q_G = 1000$, where the subscripts $E$ and $G$ represent Young's modulus and shear modulus, respectively. In all the simulations, the materials used are low-carbon steel with the following material parameters: density $\rho = 7900\ kg/m^3$, Young's modulus $E = 190 * (1 + j/Q_E)\ GPa$ and shear modulus $G = 74.2 * (1 + j/Q_G)\ GPa$ where $j = \sqrt{-1}$. The piezoelectric strips were made using PZT-4 with the same size of $8 \times 4 \times 2\ mm^3$.

### 3.2 Selective mode excitation on different structures

Compared with the existing excitation methods, the proposed in-plane excitation has many advantages. For example, it is convenient to excite the torsional modes of a shaft without additional brackets and excite vibration modes with actuators located on the nodal lines. To show the capability of the proposed excitation method, FE simulations were conducted on different structures, i.e., a rectangular bar, a cylinder, and a circular plate. For simplicity, only the results of the rectangular bar are discussed in detail and the results of the cylinder and the plate are briefly presented.

(1) Rectangular bar

Here a $120 \times 20 \times 30\ mm^3$ rectangular bar was studied. The Cartesian coordinate shown in Fig. 1 is used for mode description hereafter. Firstly, the effect of the bonded piezoelectric strips on the vibration was evaluated. The natural frequencies of the bar with or without bonded piezoelectric strips were calculated and the results are given in Tab. 1. The letters B, L, and T represent the bending modes, longitudinal modes, and torsional modes, respectively. The number after the letter denotes the mode order. This denotation will be used hereafter without specification. It can be seen from Tab. 1 that the frequency shifts caused by the additional piezoelectric strips were less than 0.5 % which is sufficiently small and can be neglected in practical engineering.

Tab. 1 Natural frequencies of the steel bar with or without bonded piezoelectric strips

| Mode | Item | Natural Frequencies of Different Orders (Hz) | | | |
|---|---|---|---|---|---|
| | | 1st | 2nd | 3rd | 4th |
| B | pure bar | 8,833.6 | 19,703 | 31,758 | 43,019 |
| | with PZTs | 8,805.7 | 19,635 | 31,631 | 42,820 |
| | errors | **0.32%** | **0.35%** | **0.40%** | **0.46%** |
| T | pure bar | 10,854 | 21,686 | 32,485 | 43,226 |
| | with PZTs | 10,809 | 21,594 | 32,342 | 43,037 |
| | errors | **0.41%** | **0.42%** | **0.44%** | **0.44%** |
| L | pure bar | 20,372 | 40,308 | - | - |
| | with PZTs | 20,339 | 40,248 | - | - |
| | errors | **0.16%** | **0.15%** | - | - |

(Note: The errors were calculated by $(I_{up} - I_{down})/I_{down} \times 100\%$, where $I_{up}$ and $I_{down}$ are the upper and lower items in the tables. In this table, $I_{up}$ is the natural frequency of a pure bar and $I_{down}$ is the natural frequency of a bar with 2 PZTs. This calculation method will be used in all the tables.)

Secondly, the mode selectivity of the proposed excitation method was validated. In traditional excitation methods, a single vibration mode is usually excited at or near the resonance frequency of this mode. If the resonance frequencies of multiple modes are close to each other, these modes would be difficult to separate and identify. In comparison, the proposed excitation method can selectively excite desired modes over a wide frequency range.

To prove this, the configurations shown in Fig. 1(b-d) were simulated and the frequency responses of the bar were monitored. Strain energies representing different vibration modes are extracted from the simulated strain and stress results. The strain distributed on the cross sections, which are parallel with the $xy$ plane, can be divided into different parts representing different vibration modes based on the symmetry features. For bending modes along the $y$-axis, the normal strain $\varepsilon_{zz}$ in the cross-section should be symmetric along the $x$-axis and antisymmetric along the $y$-axis. For longitudinal modes along the $z$-axis, the normal strain $\varepsilon_{zz}$ in the cross-section should be symmetric along both the $x$-axis and the $z$-axis. As to the torsional modes around the $z$-axis, the shear strain $\varepsilon_{xz}$ and $\varepsilon_{yz}$ in the cross-section should be antisymmetric along both the $x$-axis and $y$-axis.

The strain energies of different vibration modes in the bar can be expressed as:

$$U^{(B)} = \frac{1}{T}\int_0^T \iiint \frac{1}{2}\left(\sigma_{zz}^{(B)}\varepsilon_{zz}^{(B)}\right) dV\, dt$$
$$U^{(L)} = \frac{1}{T}\int_0^T \iiint \frac{1}{2}\left(\sigma_{zz}^{(L)}\varepsilon_{zz}^{(L)}\right) dV\, dt \quad (4)$$
$$U^{(T)} = \frac{1}{T}\int_0^T \iiint \frac{1}{2}\left(2\sigma_{xz}^{(T)}\varepsilon_{xz}^{(T)} + 2\sigma_{yz}^{(T)}\varepsilon_{yz}^{(T)}\right) dV\, dt$$

where the superscripts indicate they are part of the stress (or strain) that meet the symmetry of the corresponding vibration mode.

Using Eq. (4), the strain energy frequency responses extracted from the three simulations, in which the piezoelectric strips were arranged as shown in Fig. 1(b-d), are plotted in Fig. 3(b-d), respectively. For comparison, the strain energy frequency responses of the same bar excited by a point force $F$ at a corner point were also calculated and presented in Fig. 3(a). The point force F was set as $F_x = F_y = F_z = 1\ mN$.

As shown in Fig. 3(a), the point-force will excite the three types of vibration modes with the magnitude on the same order which brings great difficulties to mode separation and identification. In contrast, the proposed method can selectively excite the desired modes with all the other modes greatly suppressed over all the exciting frequencies, as shown in Fig. 3 (b-d). It can be seen that when the PZT strips are bonded as in Fig. 1(b), only the torsional modes are excited; as in Fig. 1(c), only the bending modes are excited and as in Fig. 1(d), only the longitudinal modes are excited. Moreover, even the 4th $B_y$ mode and the 4th T mode, whose resonance frequencies are very close, can be selectively excited using the proposed method, as seen in Fig. 3(b) and Fig. 3(c).

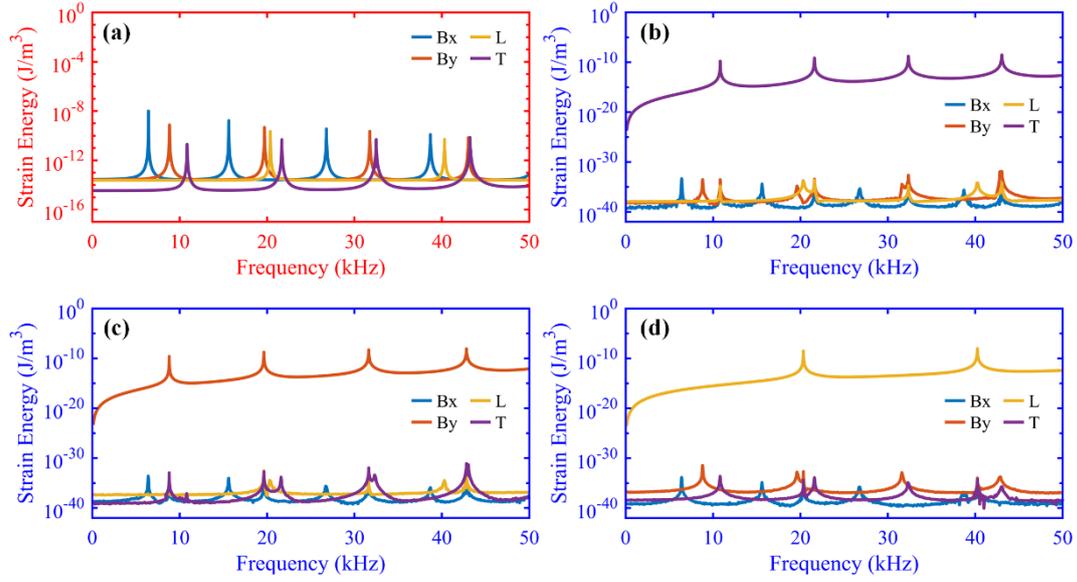

Fig. 3 The simulated strain energy frequency responses of different vibration modes calculated using Eq. (4). In (a) the bar was excited by a point force and in (b-d) the bar was excited by PZT configurations shown in Fig. 1(b-d), respectively. The subscript $x/y$ in $B_{x/y}$ denotes the bending modes along the x/y-axis.

Thirdly, the effectiveness of EMA by measuring the impedance of the PZT-structure system was validated. The simulated conductance, which represents the real part of the admittance, was monitored and plotted in Fig. 4(a). The displacement components $u_x/u_y/u_z$, which reveal the magnitude of T/B/L modes respectively, were monitored at the point $M$ shown in Fig. 1(a) and plotted in Fig. 4(b). It can be seen that good agreement is achieved between the displacement-based results and the conductance-based results. Therefore, compared with traditional MEA where extra sensors are required, in the proposed excitation method, it is more convenient to extract the vibration mode just by measuring the impedance curve, which is very quick, stable, and repeatable.

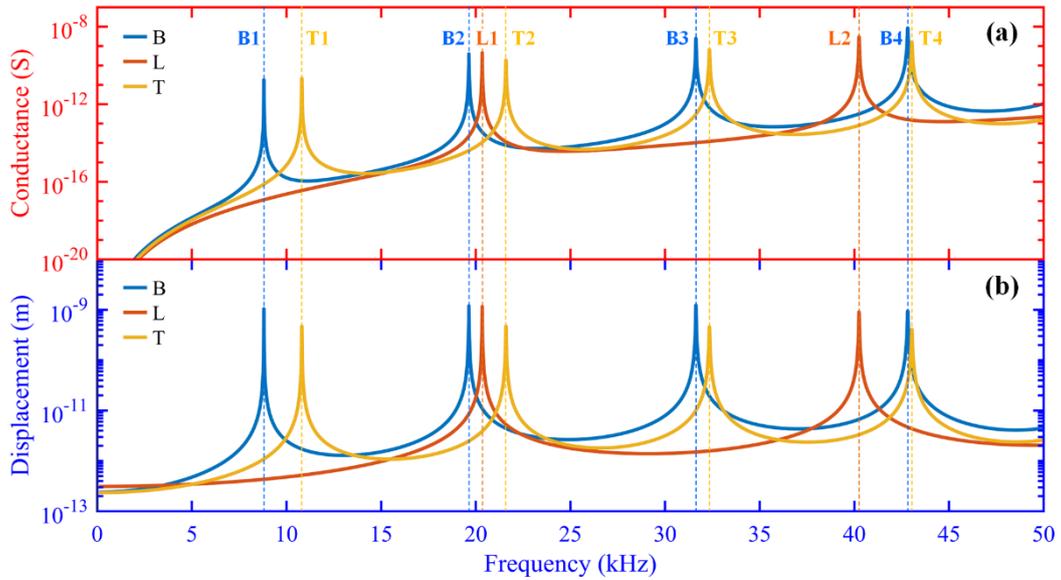

Fig. 4 Simulated conductance (a) and displacement (b) frequency response spectrum of the steel bar under different excitation layouts shown in Fig. 1(b-d).

Furthermore, from the displacement curve and the susceptance curve, the quality factors of different vibration modes can be calculated and the results are given in Table 2. Good agreement can be seen between the quality factors obtained from the displacement and the susceptance curves, and more information can be extracted. Firstly, the $Q_T$ (quality factors of the torsional modes) is mostly determined by $Q_G$ and slightly affected by $Q_E$. Secondly, the $Q_B$ and $Q_L$ (quality factors of the bending and the torsional modes) are mainly determined by $Q_E$ but also affect by $Q_G$.

Tab. 2 Simulated Quality Factors of Different Vibration Modes of a Rectangular Bar

| Mode | Item | Quality Factors of Different Orders | | | |
| --- | --- | --- | --- | --- | --- |
| | | 1st | 2nd | 3rd | 4th |
| B | $Q_{susceptance}$ | 1,818.7 | 1,591.3 | 1,448.5 | 1,289.5 |
| | $Q_{displacement}$ | 1,818.7 | 1,591.3 | 1,448.5 | 1,289.9 |
| | errors | 0.00% | 0.00% | 0.00% | -0.03% |
| T | $Q_{susceptance}$ | 1,001.7 | 1,009.7 | 1,024.5 | 1,046.3 |
| | $Q_{displacement}$ | 1,002.6 | 1,009.7 | 1,024.5 | 1,046.3 |
| | errors | -0.09% | 0.00% | 0.00% | 0.00% |
| L | $Q_{susceptance}$ | 1,896.8 | 1,619.9 | - | - |
| | $Q_{displacement}$ | 1,898.6 | 1,619.9 | - | - |
| | errors | -0.09% | 0.00% | - | - |

(2) Cylinder

A 200 mm long, 40 mm diameter cylinder was simulated with two $8 \times 2 \times 2$ mm$^3$ PZT-4 strips bonded at one end, as shown in Fig. 5(a). The conductance frequency response and the corresponding mode shape results were given in Fig. 5(b). It can be seen from the mode shapes that the excited vibration modes are the first three torsional modes. The torsional vibration characteristics of shafting structures can therefore be tested using this method without the aid of brackets.

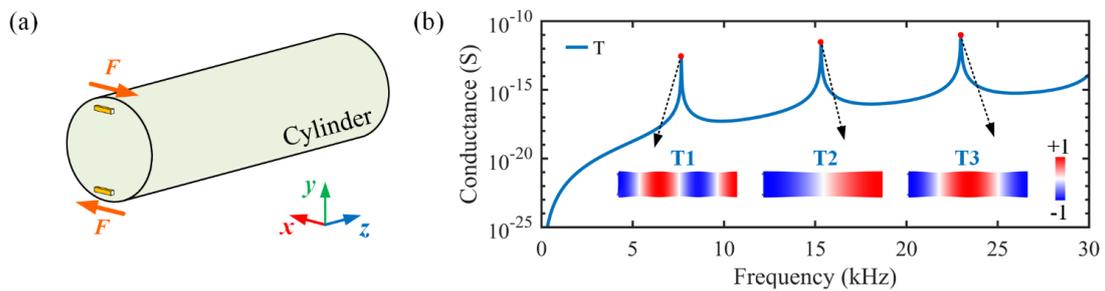

Fig. 5 Torsional excitation of a cylinder. (a) Schematic layout; (b) The conductance frequency response and the corresponding mode shape results. The mode shapes are plotted using the normalized tangential displacement.

(3) Circular plate

A 2 mm thick, 200 mm diameter plate was further simulated with six $8 \times 2 \times 2$ mm$^3$ PZT-4 strips bonded on the upper and lower faces, as shown in Fig. 6(a). The six strips were separated into three pairs and each pair generates torque around the nodal line. The simulated conductance frequency response and the corresponding mode shape results are shown in Fig. 6(b). The eigenmodes of the plate (without piezoelectric strips) over 0.1~2 kHz are also given in Fig. 6(c). It can be seen from Fig. 6(b-c) that only the 3rd bending modes are selectively excited. Note that conventional excitation method can't excite a vibration mode if the actuators are placed on nodal points or nodal lines.

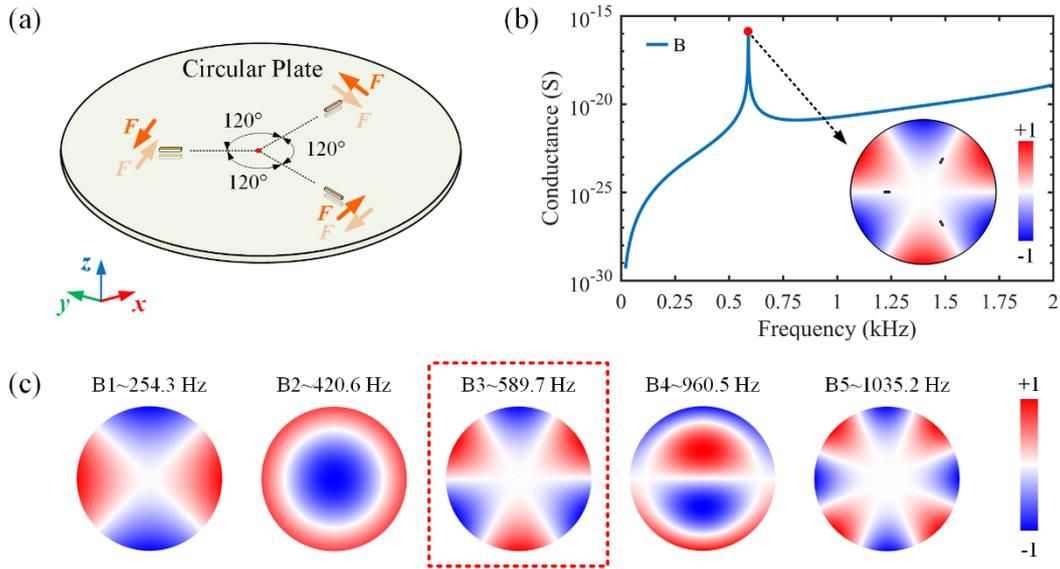

Fig. 6 Selective mode excitation of a circular plate. (a) Schematic layout; (b) The conductance frequency response and the corresponding mode shape results; (c) The first five flexural modes of the circular plate (without piezoelectric strips). The mode shapes are plotted using the out-of-plane displacement.

## 4. Experiments

To further validate the proposed excitation method, experiments were conducted on the rectangular bar and the testing setups are shown in Fig. 7. The sizes of the steel bar and the piezoelectric strips used in the experiments are the same as that in the FE simulations. The piezoelectric strips were bonded onto the bar using the 502 epoxy adhesive.

The electromechanical impedance of the PZT-structure composite system was measured by a high-resolution impedance analyzer (IM3570, HIOKI, Japan), as shown in Fig. 7(a). A 2D laser Doppler vibrometer was used to monitor the bar's corner point velocity, as shown in Fig. 7(b). The piezoelectric strips were driven by a continuous sine electric signal generated by a function generator (33220A, Agilent, USA) and amplified by a power amplifier (ATA-2021H, Aigtek, China). It can be seen from Fig. 7 that the testing setup required for the impedance-based mode analysis is much simpler than that for the velocity-based one.

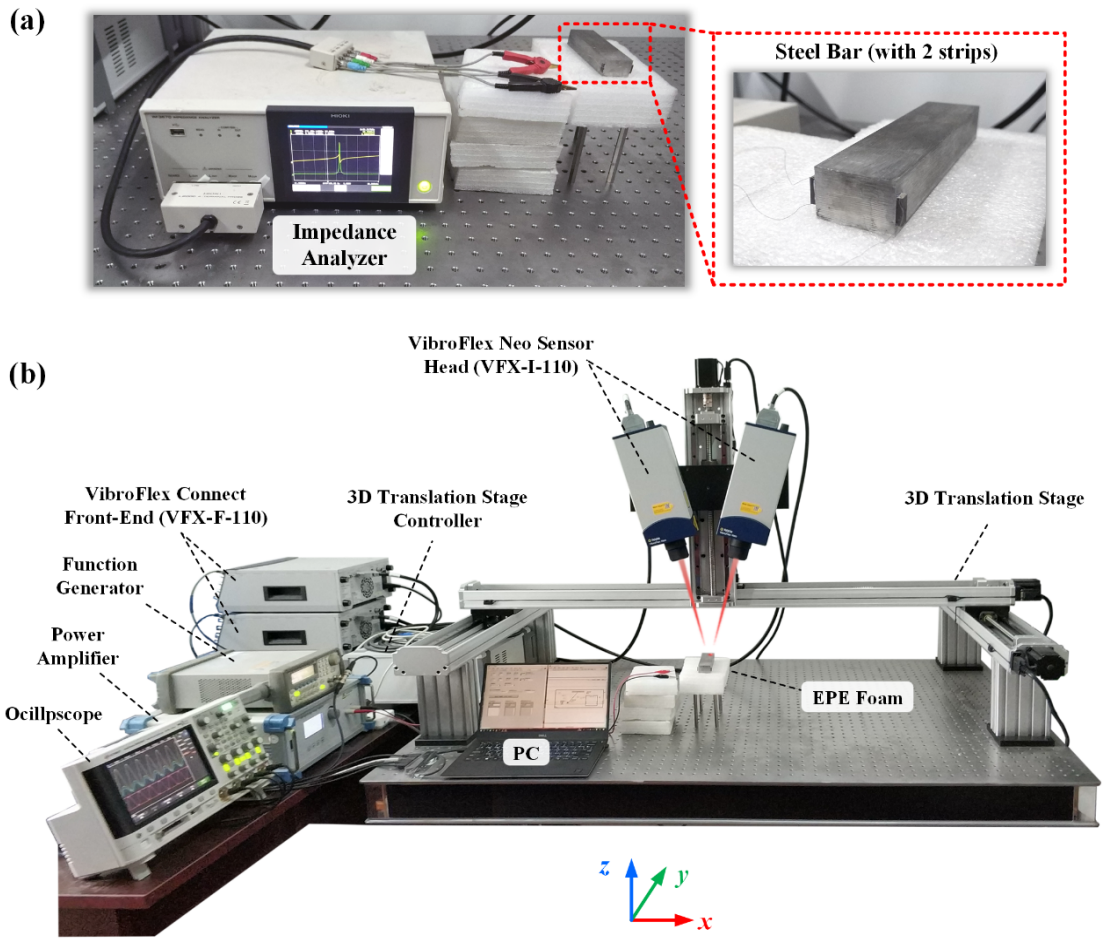

Fig. 7 Experimental setups to examine the performance of the proposed excitation method in EMA. (a) Impedance-based setup; (b) Velocity-based setup.

Both the measured velocity and the conductance results were plotted in Fig. 8. The differences between the natural frequencies calculated from the velocity and the conductance are rather small (< 0.02%) and are thus negligible. The quality factors extracted from both methods are given in Tab. 3. Unlike the simulated results in which very good agreement is obtained, some differences can be seen between the quality factors extracted from the experimental velocity and the conductance curve. This should be caused by the use of different driving power during the velocity-based and conductance-based experiments. The adhesive can also have some influences on the measurement results. Note all the discrepancies are within ±8 %, they are acceptable since the quality factor of a structure depends on many factors and is not a constant.

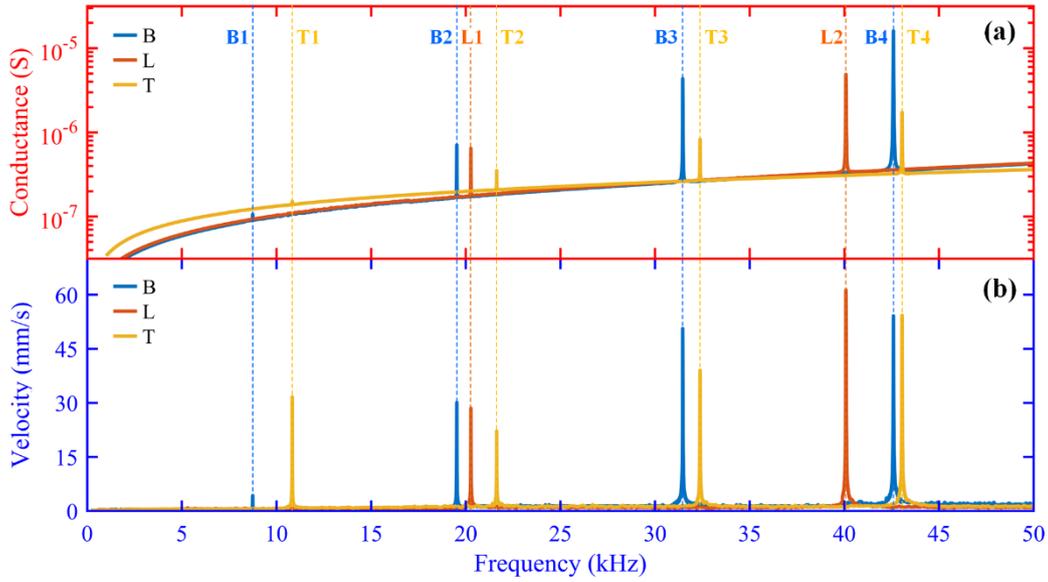

Fig. 8 Experimental frequency response spectrum of the conductance (a) and the velocity (b) of a rectangular bar under different excitation layouts.

Tab. 3 Measured Quality Factors of Different Vibration Modes from Velocity and Impedance

| Mode | Item | Quality Factors of Different Orders (Hz) | | | |
|---|---|---|---|---|---|
| | | 1st | 2nd | 3rd | 4th |
| B | $Q_{velocity}$ | 1,474.3 | 2,040.5 | 2,162.1 | 1,997.6 |
| | $Q_{susceptance}$ | 1,408.4 | 2,064.8 | 2,182.4 | 2,127.9 |
| | errors | **4.47%** | **-1.19%** | **-0.94%** | **-6.52%** |
| T | $Q_{velocity}$ | 1,645.8 | 2,036.7 | 2,248.2 | 2,143.7 |
| | $Q_{susceptance}$ | 1,514.4 | 2,025.4 | 2,174.7 | 2,178.6 |
| | errors | **7.98%** | **0.55%** | **3.27%** | **-1.63%** |
| L | $Q_{velocity}$ | 1,922.9 | 2,105.0 | - | - |
| | $Q_{susceptance}$ | 1,824.0 | 2,004.1 | - | - |
| | errors | **5.14%** | **4.79%** | - | - |

To verify if the excited vibration modes are the desired ones, the bar's velocity field was scanned using the 2D laser Doppler vibrometer with the aid of a 3D translation stage. The y-component velocity field of the first four bending modes were shown in Fig. 9. It can be seen from the nodal lines that the excited modes are truly the bending modes. So, the bending, torsional, and longitudinal vibration modes had been successfully excited by using the proposed method.

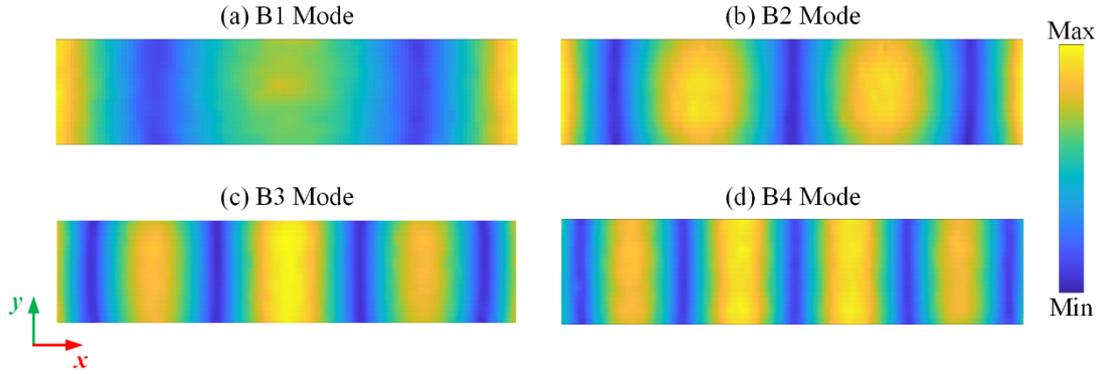

Fig. 9 The y-component velocity field of the first four excited bending modes of a rectangular bar scanned by a 2D laser Doppler vibrometer.

## 5. Conclusions

In summary, an in-plane vibration excitation method was proposed for EMA based on the thickness-shear ($d_{15}$) piezoelectric transducers. Both simulations and experiments showed that the proposed method can selectively excite the desired modes in a wide frequency range. Unlike the conventional velocity, acceleration, or strain based EMA, in the proposed method, the electromechanical impedance of the system was measured from which both the natural frequency and quality factor can be extracted. Only an impedance analyzer is required to conduct the measurement, which considerably reduced the experimental complexity. Due to its simplicity and flexibility, the proposed excitation method is expected to be widely used in near future.

### Acknowledgements

This work is supported by the National Natural Science Foundation of China under Grant No. 12172007.